\documentclass[
    reprint,
    amsmath,
    amssymb,
    aps,
    superscriptaddress,
    nofootinbib
]{revtex4-1}

\usepackage{array}[=2016-10-06]
\usepackage{dcolumn}
\usepackage{tabularx}
\usepackage{booktabs}

\usepackage[utf8]{inputenc}
\usepackage{graphicx}
\usepackage{bm}
\usepackage{float}
\usepackage{amsmath,amsthm,amsfonts,amssymb}
\usepackage[svgnames]{xcolor}
\usepackage{mathtools}
\usepackage{caption}
\usepackage{subcaption}
\usepackage{soul}
\usepackage{appendix}
\usepackage{orcidlink}
\usepackage{comment}
\usepackage{hyperref}

\definecolor{teal}{RGB}{0,128,128}

\hypersetup{
    colorlinks=true,
    linkcolor=teal,
    citecolor=teal,
    urlcolor=teal
}

\begin{document}

\title{Charged black string immersed in a quintessence fluid and string cloud}

\author{Leonardo G. Barbosa \orcidlink{0009-0007-3468-3718}}
\email{leonardo.barbosa@posgrad.ufsc.br}
\affiliation{Departamento de F\'isica, CFM - Universidade Federal de Santa Catarina, \\ Caixa Postal 5064, CEP 880.35-972, Florian\'opolis, SC, Brazil.}

\author{Franciele M. da Silva \orcidlink{0000-0003-2568-2901}} 
\email{franciele.m.s@ufsc.br}
\affiliation{Departamento de F\'isica, CFM - Universidade Federal de Santa Catarina, \\ Caixa Postal 5064, CEP 880.35-972, Florian\'opolis, SC, Brazil.}
\affiliation{Theoretical Astrophysics, Institute for Astronomy and Astrophysics, University of T\"{u}bingen, 72076 T\"{u}bingen, Germany}

\begin{abstract}
	We present a new static solution describing a charged black string immersed in a Kiselev-type quintessence fluid and a string cloud. The metric and field equations are solved for a general quintessence state parameter, with explicit results provided for the physically relevant case $w_q = -2/3$. We analyze the event-horizon structure and the Kretschmann scalar, verify energy-condition constraints, and derive thermodynamic properties including the Hawking temperature and heat capacity and the Helmholtz free energy. Finally, we investigate the existence of an exterior photon cylinder for circular null geodesics, finding that, under the parameter assumptions adopted here, no such cylinder occurs outside the outer horizon. The solution generalizes known charged black-string spacetimes by simultaneously including quintessence and a string-cloud parameter.
\end{abstract}

\maketitle

\section{Introduction}\label{Introduction}

The gravitational collapse of a massive star is commonly associated with the formation of a black hole in a highly compact region, a picture often summarized by the hoop conjecture~\cite{Thorne1972mwm}. In its standard form, this argument suggests that black hole formation is closely tied to the concentration of matter within a sufficiently small spatial region. Nevertheless, this viewpoint has long been discussed in the context of gravitational collapse and cosmic censorship. This interpretation relies on the assumption of a spacetime with a vanishing cosmological constant. When a negative cosmological constant is introduced, the global structure of the spacetime changes in an essential way, and the usual geometric restrictions on black hole horizons are no longer as strict. Within this broader setting, black strings emerge as an important family of exact solutions to the Einstein field equations~\cite{Lemos:1994xp}. These configurations were originally proposed as cylindrical counterparts of anti-de Sitter black holes. Additionally, they are characterized by horizons that extend along one spatial direction rather than closing into the usual spherical shape. As a result, they provide a useful example of how non-spherical horizon geometries can arise in spacetimes with a negative cosmological constant. The static solutions were later extended to include charge and rotation, further enriching the class of known black string geometries~\cite{Lemos:1995cm}.

Black strings have also played an important role in the study of thermodynamics, conserved charges, and semiclassical emission. Their fundamental thermodynamic properties and entropy limits were analyzed in detail in the early literature~\cite{Bekenstein:1973ur, Gibbons:1977mu}. Later, the same framework was expanded to include extended phase spaces~\cite{Cai:2013qga} and charged torus-like configurations~\cite{Ditta:2023gin}. In parallel, Hawking radiation~\cite{Hawking:1975vcx} from black strings has been investigated through tunneling approaches~\cite{Parikh:1999mf} and related semiclassical methods~\cite{RamonMedrano:1999in, Gambini:2013nea}, with results obtained for scalar, fermionic, and vector particles. Furthermore, geometries with cylindrical symmetry provide a critical arena for examining dynamic and thermodynamic stability. In particular, when it regards to the Gregory-Laflamme instability of black strings and the semiclassical instabilities of higher-dimensional rotating black holes~\cite{Gregory:1993vy, Gregory:1994bj, Dias:2009iu, Monteiro:2009ke, Dias:2010eu, Yoo:2011vu}.

Recent studies have extended both neutral, charged, and regular black string spacetimes by coupling them to background matter fields. Some of these studies include modified gravity models such as $f(R,T)$ theory~\cite{Santos:2026bjq}, Rainbow Gravity~\cite{Darlla:2023qgf}, dynamical Chern--Simons modified gravity~\cite{Cisterna:2018jsx}. In parallel, other studies have investigated regular black strings~\cite{Jusufi:2022rbt}, as well as anisotropic Kiselev fluids~\cite{Kiselev:2002dx, Ghosh:2015ovj, Toshmatov:2015npp}. The Kiselev fluid, characterized by an equation-of-state parameter $w_q$, provides an effective description of quintessence-like dark energy~\cite{Caldwell:1997ii, Steinhardt:2003st, Copeland:2006wr} and other anisotropic distributions, interpolating between different known solutions (e.g., an effective cosmological constant for $w_q=-1$). Such fluids have been successfully incorporated into black string geometries~\cite{Ali:2019mxs}. Beyond quintessence, the spacetime around a black string may also be affected by a string cloud~\cite{Letelier:1979ej}. The coexistence of these matter sources modifies the asymptotic structure, the behavior of singularities, and the thermodynamic properties of the spacetime~\cite{deMToledo:2018tjq, Cunha:2022kep, Ahmed:2025sav, Simao:2025xhd}.

Several particular configurations of black strings surrounded by anisotropic matter have been investigated in the literature. Black strings surrounded by a perfect fluid of dark matter (PFDM) have been studied~\cite{Barbosa:2026guu}, as well as charged black strings surrounded by a Kiselev anisotropic fluid~\cite{Barbosa:2025scy}. Neutral black strings surrounded by a Kiselev anisotropic fluid and a string cloud have also been considered~\cite{Deglmann:2025mcl}. However, the combination of a charged black string within Einstein--Maxwell theory, a Kiselev anisotropic fluid, and a string cloud has not yet been investigated in the literature, to the best of our knowledge.

In this work, we solve the Einstein-Maxwell field equations for a charged black string coupled to a Kiselev anisotropic fluid and string cloud matter, obtaining an exact spacetime solution that incorporates the combined effects of these sources. We then examine in detail the resulting geometry, with emphasis on how the matter content modifies the singularity structure, the location and nature of the event horizon, and the asymptotic behavior of the spacetime. We also analyze the relevant energy conditions to assess the physical viability of the solution and study the Hawking temperature in order to understand its thermodynamic properties. Finally, we investigate the geodesic motion associated with this background, which provides further insight into the dynamical features and observable consequences of the spacetime.

This paper is organized as follows. In Sec.~\ref{Field_equation} we present the field equations for the matter distributions considered and obtain a new charged black-string solution for a general quintessence state parameter $w_q$. While the general form is given, our subsequent analysis focuses explicitly on the case $w_q=-2/3$. In Sec.~\ref{Energy_conditions} we analyze the weak and strong energy conditions and discuss their implications for this solution. In Sec.~\ref{Event_horizon_and_singularity_structure} we study the corresponding event horizon, including the extremal configuration, and perform an analysis of the Kretschmann scalar. In Sec.~\ref{Thermodynamics} we compute the Hawking temperature and the heat capacity, discussing the associated critical radius, and complement this local analysis with the Helmholtz free energy to assess global thermodynamic stability. In Sec.~\ref{Photon_cylinder} we investigate the existence of an exterior photon cylinder for circular null geodesics, showing that, under the parameter assumptions adopted here, no such cylinder is found outside the outer horizon. Finally, in Sec.~\ref{Discussion_and_conclusions} we present our conclusions.

\section{Field equations}\label{Field_equation}

In this section, we obtain a static charged black-string solution in the presence of a Kiselev fluid and a string cloud. We consider Einstein's field equations with a negative cosmological constant, adopting $G=c=1$ and the signature $(-,+,+,+)$:

\begin{equation}
	R_{\mu\nu}-\frac{1}{2}Rg_{\mu\nu}+\Lambda g_{\mu\nu}
	=8\pi T_{\mu\nu},
\end{equation}
where $R_{\mu\nu}$ is the Ricci tensor, $R$ is the curvature scalar, $g_{\mu\nu}$ is the metric tensor, and $T_{\mu\nu}$ is the energy-momentum tensor. We write the cosmological constant as $\Lambda=-3/\ell^2$, where $\ell$ is the AdS radius. The matter distribution consists of three non-interacting sources,

\begin{equation}
	T_{\mu\nu}
	=T_{\mu\nu}^{(1)}+T_{\mu\nu}^{(2)}+T_{\mu\nu}^{(3)}.
\end{equation}

The first contribution is given by the electromagnetic energy-momentum tensor,
\begin{equation}
	T_{\mu\nu}^{(1)}
	=\frac{1}{4\pi}\left(
	F_{\mu}{}^{\eta}F_{\nu\eta}
	-\frac{1}{4}g_{\mu\nu}F_{\alpha\beta}F^{\alpha\beta}
	\right),
\end{equation}
where $F_{\mu\nu}=\nabla_\mu A_\nu-\nabla_\nu A_\mu$ is the electromagnetic field tensor, $A_\mu$ is the four-potential, and $\nabla_\mu$ denotes the covariant derivative. We take $A_\mu=-h(r)\delta_\mu^t$, with $h(r)$ determined by Maxwell's equation,

\begin{equation}\label{MaxwellI}
	\nabla_\mu F^{\mu\nu}=0.
\end{equation}

The second contribution $T_{\mu\nu}^{(2)}$ is taken as the anisotropic energy-momentum tensor introduced by Kiselev to model quintessence~\cite{Kiselev:2002dx}. In coordinates adapted to the string geometry, the nonzero components are
\begin{align}
	T_t{}^{(2)t}
	&=T_r{}^{(2)r}=-\rho_q,\\
	T_\varphi{}^{(2)\varphi}
	&=T_z{}^{(2)z}=\tfrac{1}{2}\rho_q(3w_q+1),
\end{align}
where $w_q$ is the equation-of-state parameter and $\rho_q$ is the energy density. The transverse pressure can be written as
\begin{equation}
	p_q=\tfrac{1}{2}\rho_q(3w_q+1),
	\qquad -1\leq w_q\leq-\tfrac{1}{3}.
\end{equation}

The parameter $w_q$ determines the relation between the energy density and the transverse pressure of the Kiselev fluid.

The third contribution to the matter sector is given by the energy-momentum tensor of a string cloud~\cite{Letelier:1979ej},
\begin{align}
	T_t{}^{(3)t}
	&=T_r{}^{(3)r}=-\frac{a}{r^2},\\
	T_\varphi{}^{(3)\varphi}
	&=T_z{}^{(3)z}=0,
\end{align}
where $a$ is a positive real parameter. For a static geometry with cylindrical symmetry, we consider the line element
\begin{equation}
	ds^2=-f(r)dt^2+\frac{dr^2}{f(r)}
	+r^2d\varphi^2+\frac{r^2}{\ell^2}dz^2,
\end{equation}
where $-\infty<t<+\infty$, $0\leq r<+\infty$, $0\leq\varphi<2\pi$, and $-\infty<z<+\infty$. Substituting the metric and the matter sources into Einstein's equations gives
\begin{align}
	\frac{f'(r)}{r}+\frac{f(r)}{r^2}-\frac{3}{\ell^2}
	&=-[h'(r)]^2-8\pi\rho_q-\frac{8\pi a}{r^2},
	\label{eq.1}\\
	\frac{f''(r)}{2}+\frac{f'(r)}{r}-\frac{3}{\ell^2}
	&=[h'(r)]^2+4\pi\rho_q(3w_q+1).
	\label{eq.2}
\end{align}

\subsection{Solution}

Integrating Maxwell's equation yields $h'(r)=Q/r^2$, where $Q$ is an integration constant associated with the electric charge. Substituting this expression into Eqs.~\eqref{eq.1} and~\eqref{eq.2} and eliminating $\rho_q$, we obtain

\begin{equation}\label{eq.3}
	\begin{split}
		&f''(r)+3(w_q+1)\frac{f'(r)}{r}\\
		&+(3w_q+1)\frac{f(r)}{r^2}
		+(3w_q-1)\frac{Q^2}{r^4}\\
		&-9\ell^{-2}(w_q+1)
		+\frac{8\pi a}{r^2}(3w_q+1)=0.
	\end{split}
\end{equation}

Direct integration of this second-order differential equation gives

\begin{equation}\label{Solution}
	f(r)=\frac{r^2}{\ell^2}-\frac{2M}{r}
	+\frac{Q^2}{r^2}
	-\frac{N_q}{r^{3w_q+1}}-\alpha,
\end{equation}

where $M$ and $Q$ are the mass and charge parameters, respectively. The constant $N_q$ is associated with the Kiselev fluid, while $\alpha=8\pi a$ characterizes the string-cloud contribution.

	The radial dependence of the Kiselev term varies with $w_q$. For $w_q=-1$, it is proportional to $r^2$ and contributes as an effective cosmological constant. For $w_q=-1/3$, it becomes a constant term with the same radial dependence as the string-cloud contribution. Here, we focus on $w_q=-2/3$, for which

\begin{equation}\label{eq:metric_special}
	f(r)=\frac{r^2}{\ell^2}-\frac{2M}{r}
	+\frac{Q^2}{r^2}-N_qr-\alpha.
\end{equation}

	The Kiselev fluid therefore contributes a term linear in $r$, while the string cloud introduces a constant shift. The mass and charge contributions are proportional to $r^{-1}$ and $r^{-2}$, respectively, and the leading large-radius term is $r^2/\ell^2$.
	
	Table~\ref{tab:static_comparison}, presented in Appendix~\ref{app:comparison}, compares this metric with related static black-string solutions. The limits $\alpha=0$ and $Q=0$ recover the metric functions of Refs.~\cite{Barbosa:2025scy,Deglmann:2025mcl}, respectively.

	To illustrate the dependence on the string-cloud parameter, we consider the radial profile of $f(r)$ for $w_q=-2/3$ and different values of $\alpha$, keeping $\ell$, $M$, $Q$, and $N_q$ fixed. Since the string-cloud contribution is $-\alpha$, increasing $\alpha$ shifts the curve downward by a constant amount.

\begin{figure}[h!]
	\centering
	\includegraphics[width=\columnwidth]{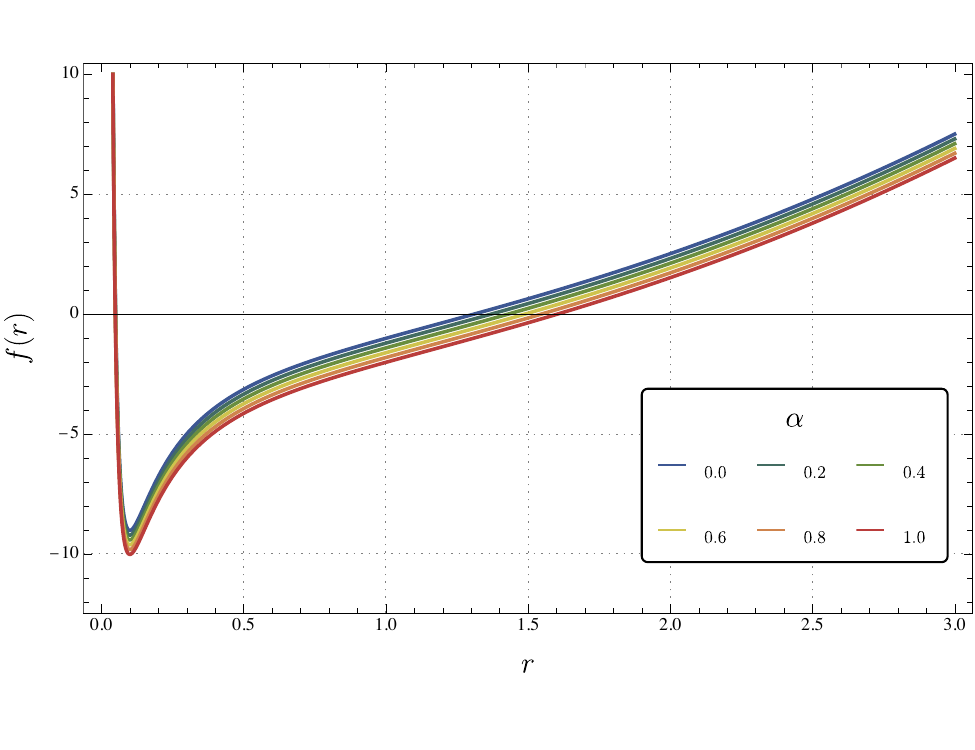}
	\caption{Metric function $f(r)$ as a function of $r$ for
		$w_q=-2/3$, with $\ell=1$, $M=0.9$, $Q=0.3$, and
		$N_q=0.3$. The curves correspond to
		$\alpha\in\{0.0, 0.2, 0.4, 0.6, 0.8, 1.0\}$.}
	\label{fig:metric_function}
\end{figure}

\begin{table}[h!]
	\caption{
			Inner and outer horizon radii for the parameter values
			specified in Fig.~\ref{fig:metric_function}.
	}
	\begin{center}
		\begin{tabular}{c c c}
			\hline
			\hline
			$\alpha$ & $r_h^{\rm int}$ & $r_h^{\rm ext}$\\
			\hline
			0.0 & 0.0500 & 1.3095 \\
			0.2 & 0.0497 & 1.3700 \\
			0.4 & 0.0494 & 1.4297 \\
			0.6 & 0.0492 & 1.4884 \\
			0.8 & 0.0489 & 1.5461 \\
			1.0 & 0.0487 & 1.6028 \\
			\hline
			\hline
		\end{tabular}
		\label{T}
	\end{center}
\end{table}

	For the parameter values considered, the equation $f(r)=0$ has two positive roots, corresponding to the inner and outer horizons. Their numerical values are listed in Table~\ref{T}. Over this range of $\alpha$, the inner horizon radius decreases, while the outer horizon radius increases.

\section{Energy conditions}\label{Energy_conditions}

In this section, we analyze the energy conditions for the black string solution obtained in the previous section. We consider the weak and strong energy conditions, taking into account the contributions of the Kiselev fluid, electromagnetic field, string cloud, and cosmological constant. Substituting the metric function into Eqs.~(\ref{eq.1}) and (\ref{eq.2}), the quintessence energy density $\rho_q$ and the transverse pressure $p_q$ are given by
\begin{align}
	\rho_q &=-\frac{3N_qw_q}{8\pi r^{3(w_q+1)}},\\
	p_q &=-\frac{3N_qw_q(3w_q+1)}{16\pi r^{3(w_q+1)}}.
\end{align}

The integration constant $N_q$ therefore controls the normalization and sign of the Kiselev energy density. Since $w_q\in[-1,-1/3]$, one has $w_q<0$, so that $N_q>0$ leads to $\rho_q>0$, whereas $N_q<0$ leads to $\rho_q<0$. Hence, requiring a positive quintessence energy density selects $N_q>0$ within the considered range of $w_q$.

To distinguish the contribution of each source, Table~\ref{tab:energy_sectors} summarizes the corresponding energy density and principal pressures. We use $T^\mu{}_\nu=\mathrm{diag}(-\rho,p_r,p_\varphi,p_z)$, with $p_\varphi=p_z\equiv p_\perp$, for the total energy-momentum tensor. The cosmological constant, although kept on the geometric side of Einstein's equations, is included here as an effective contribution to the total energy conditions.

	\begin{table}[h!]
		\caption{Contributions of the different sectors to the energy density and principal pressures.}
		\begin{center}
			\small
			\renewcommand{\arraystretch}{1.35}
			\setlength{\tabcolsep}{3pt}
			
			\begin{tabular}{c c c c}
				\hline
				\hline
				\textbf{Sector} &
				$\boldsymbol{\rho}$ &
				$\boldsymbol{p_r}$ &
				$\boldsymbol{p_\perp}$ \\
				\hline
				
				Kiselev fluid &
				$\rho_q$ &
				$-\rho_q$ &
				$\frac{1}{2}(3w_q+1)\rho_q$ \\
				
				Maxwell field &
				$\frac{Q^2}{8\pi r^4}$ &
				$-\frac{Q^2}{8\pi r^4}$ &
				$\frac{Q^2}{8\pi r^4}$ \\
				
				String cloud &
				$\frac{a}{r^2}$ &
				$-\frac{a}{r^2}$ &
				$0$ \\
				
				Cosmological constant &
				$-\frac{3}{8\pi\ell^2}$ &
				$\frac{3}{8\pi\ell^2}$ &
				$\frac{3}{8\pi\ell^2}$ \\
				
				\hline
				\hline
			\end{tabular}
			
			\label{tab:energy_sectors}
		\end{center}
	\end{table}

	As seen from Table~\ref{tab:energy_sectors}, the radial combination satisfies $\rho+p_r=0$. The remaining combinations involve the total energy density and the transverse pressures, which enter the conditions discussed below.

\subsection{WEC}

	The weak energy condition (WEC) requires the energy density measured by timelike observers to be nonnegative. For the diagonal energy-momentum tensor considered here, it is expressed through the following inequalities:

\begin{equation}
	\rho\geq0,\quad
	\rho+p_r\geq0,\quad
	\rho+p_\varphi\geq0,\quad
	\rho+p_z\geq0.
\end{equation}

	Since $\rho+p_r=0$ and the two transverse pressures are equal, the WEC reduces to two independent inequalities. Including all contributions listed in Table~\ref{tab:energy_sectors}, we obtain

\begin{align}
	-\frac{3N_qw_q}{8\pi r^{3w_q+3}}
	+\frac{a}{r^2}
	+\frac{Q^2}{8\pi r^4}
	-\frac{3}{8\pi\ell^2}
	&\geq0,
	\label{cond1}\\
	-\frac{9(w_q+1)N_qw_q}{16\pi r^{3w_q+3}}
	+\frac{Q^2}{4\pi r^4}
	+\frac{a}{r^2}
	&\geq0.
	\label{cond2}
\end{align}

	Equation~\eqref{cond1} corresponds to $\rho\geq0$, whereas Eq.~\eqref{cond2} corresponds to $\rho+p_\perp\geq0$. The electromagnetic field and string cloud contribute positively to both expressions for $a\geq0$. The cosmological constant contributes negatively to the total energy density but cancels from the transverse combination. The Kiselev contribution depends on $N_q$ and $w_q$. Both inequalities must be considered when identifying the region where the WEC is satisfied.
	
	For the case of interest, $w_q=-2/3$, the two conditions become

\begin{align}
	\frac{N_q}{4\pi r}
	+\frac{a}{r^2}
	+\frac{Q^2}{8\pi r^4}
	-\frac{3}{8\pi\ell^2}
	&\geq0,
	\label{cond1_special}\\
	\frac{N_q}{8\pi r}
	+\frac{Q^2}{4\pi r^4}
	+\frac{a}{r^2}
	&\geq0.
	\label{cond2_special}
\end{align}

\begin{figure}[htbp]
	\centering
	\includegraphics[width=\columnwidth]{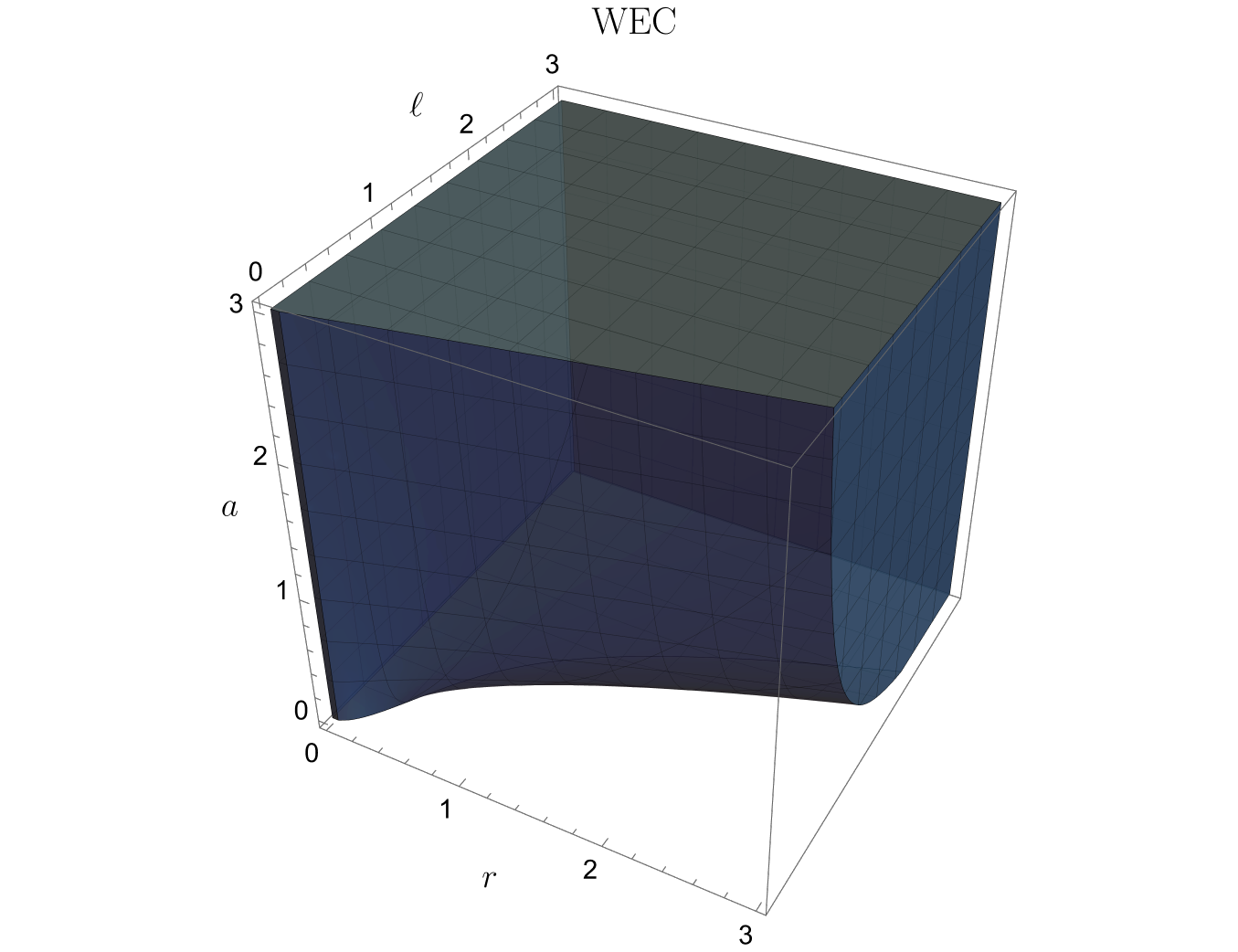}
	\caption{
			Region in the $(r,\ell,a)$ space where the weak energy
			condition is satisfied for $w_q=-2/3$, with $N_q=1$ and
			$Q=0.5$. The condition $\rho\ge0$ is dominated by the
			electromagnetic and string-cloud terms at small $r$, while
			the cosmological-constant term becomes increasingly relevant
			at large $r$.
	}
	\label{fig:wec}
\end{figure}

	The region in the $(r,\ell,a)$ space where both inequalities hold simultaneously is shown in Fig.~\ref{fig:wec}.

\subsection{SEC}

	We next consider the strong energy condition (SEC), which involves the sum of the energy density and principal pressures. In addition to the inequalities $\rho+p_i\geq0$, with $i=r,\varphi,z$, it requires

\begin{equation}
	\rho+p_r+p_\varphi+p_z\geq0.
\end{equation}

	For the sources considered here, the inequalities $\rho+p_i\geq0$ reduce to Eq.~\eqref{cond2}. Substituting the energy density and pressures from Table~\ref{tab:energy_sectors} into the remaining condition gives

\begin{equation}
	-\frac{3(3w_q+1)w_qN_q}{8\pi r^{3w_q+3}}
	+\frac{Q^2}{4\pi r^4}
	+\frac{3}{4\pi\ell^2}
	\geq0.
	\label{eq:sec}
\end{equation}

	In this expression, the string-cloud contribution cancels, while the electromagnetic field and the negative cosmological constant give positive terms. The Kiselev contribution retains its dependence on $N_q$ and $w_q$; for $w_q=-2/3$ and $N_q>0$, it enters with a negative sign. Explicitly, for $w_q=-2/3$,

\begin{equation}
	-\frac{N_q}{4\pi r}
	+\frac{Q^2}{4\pi r^4}
	+\frac{3}{4\pi\ell^2}
	\geq0.
	\label{eq:sec_special}
\end{equation}

\begin{figure}[htbp]
	\centering
	\includegraphics[width=\columnwidth]{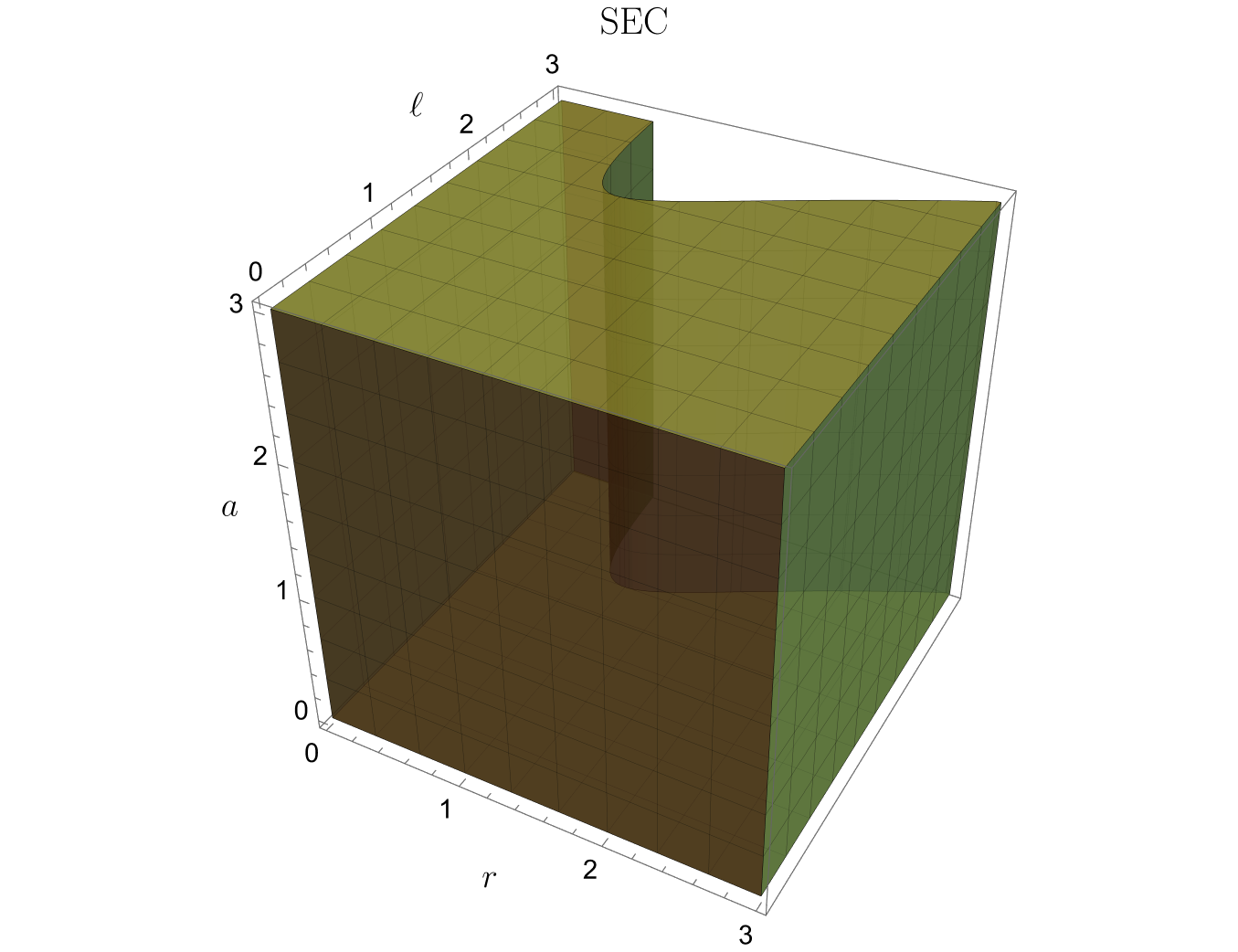}
	\caption{
			Region in the $(r,\ell,a)$ space where the strong energy
			condition is satisfied for $w_q=-2/3$, with $N_q=1$ and
			$Q=0.5$. Since the string-cloud term cancels from
			$\rho+p_r+p_\varphi+p_z$, the region is uniform along the
			$a$ direction. The boundary reflects the balance between the
			Kiselev, electromagnetic, and cosmological-constant
			contributions.
	}
	\label{fig:sec}
\end{figure}

	The SEC is evaluated by considering Eq.~\eqref{eq:sec_special} together with Eq.~\eqref{cond2_special}. The corresponding region in the $(r,\ell,a)$ space is shown in Fig.~\ref{fig:sec}. Since Eq.~\eqref{eq:sec_special} does not depend on $a$, the allowed region is uniform along the $a$ direction, and the boundary is determined only by the competition between the Kiselev, electromagnetic, and cosmological-constant terms.

\begin{figure}[htbp]
	\centering
	\includegraphics[width=\columnwidth]{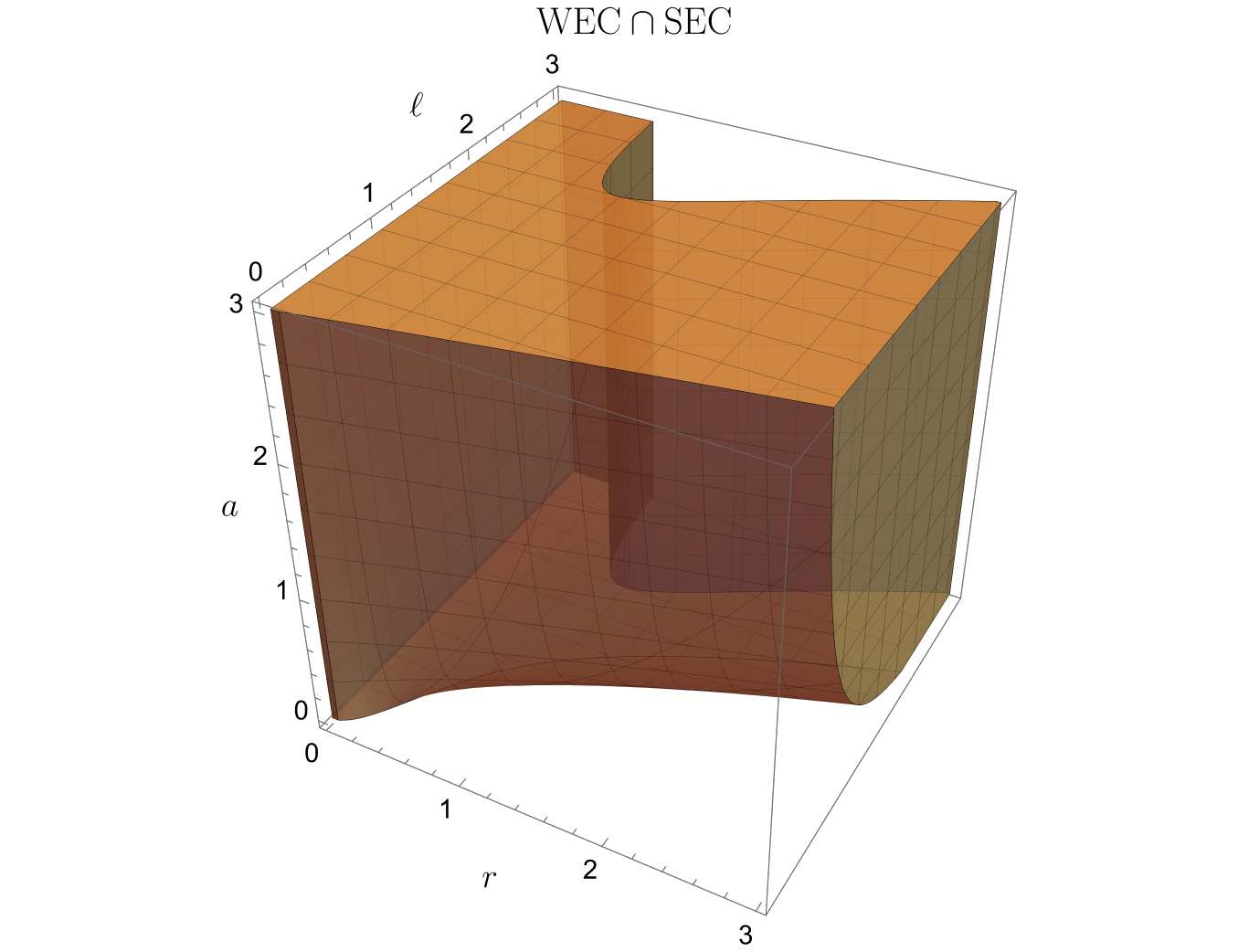}
	\caption{
			Region in the $(r,\ell,a)$ space where the weak and strong
			energy conditions are simultaneously satisfied for
			$w_q=-2/3$, with $N_q=1$ and $Q=0.5$. The region corresponds
			to the intersection of the conditions defined by
			Eqs.~\eqref{cond1}, \eqref{cond2}, and~\eqref{eq:sec}.
	}
	\label{fig:wec_sec}
\end{figure}

	The two energy conditions share the transverse inequality~\eqref{cond2}, but impose different additional restrictions through Eqs.~\eqref{cond1} and~\eqref{eq:sec}. Their simultaneous satisfaction is therefore determined by the overlap of the regions defined by these three inequalities. This overlap is shown in Fig.~\ref{fig:wec_sec}.

\section{Event horizon and singularity structure}\label{Event_horizon_and_singularity_structure}

In this section we study the event horizon solution and the structure of singularities via the Kretschmann scalar. In general, the equation that characterizes the event horizon for the solution (\ref{Solution}) is given by
\begin{equation}
	\frac{r^{2}_{h}}{\ell^{2}}-\frac{2M}{r_{h}}+\frac{Q^{2}}{r^{2}_{h}}-\frac{N_{q}}{r^{3w_{q}+1}_{h}}-\alpha=0
\end{equation}

regarding the case of interest $(w_q = -2/3)$, we can write
\begin{equation}\label{eq:horizon_quartic}
    r^{4}-N_{q}\ell^{2}r^{3}-\alpha\ell^{2}r^{2}-2M\ell^{2}r+Q^{2}\ell^{2}=0,
\end{equation}
which is a fourth-degree algebraic equation for the radius of the event horizon. Although Eq.~(\ref{eq:horizon_quartic}) admits, in principle, an exact solution via Ferrari's method, the resulting radicals are too lengthy to offer direct physical insight, so we solve it numerically for fixed $\ell$, $M$, $Q$, and $N_q$, varying the string-cloud parameter $\alpha$ (Fig.~\ref{fig:metric_function} and Table~\ref{T}). 

The horizon structure can be further characterized by analyzing the conditions for a degenerate horizon, which marks the transition between a black string with two horizons and a naked singularity. For an extremal configuration, we require $f(r_e)=0$ and $f'(r_e)=0$, which, for the solution (\ref{Solution}), yield \cite{Nam:2018ltb,Fernando:2016qhq}:
\begin{align}
r_e^4 - N_q \ell^2 r_e^3 - \alpha \ell^2 r_e^2 - 2M \ell^2 r_e + Q^2 \ell^2 &= 0, \label{eq:ext1} \\
4r_e^3 - 3N_q \ell^2 r_e^2 - 2\alpha \ell^2 r_e - 2M \ell^2 &= 0. \label{eq:ext2}
\end{align}

Eliminating $M$ from Eqs.~(\ref{eq:ext1}) and~(\ref{eq:ext2}) gives the extremal-radius condition
\begin{equation}
3r_e^4 - 2N_q \ell^2 r_e^3 - \alpha \ell^2 r_e^2 - Q^2 \ell^2 = 0, \label{eq:re}
\end{equation}
and the corresponding extremal mass
\begin{equation}
M_{\text{ext}} = \frac{2r_e^3}{\ell^2} - \frac{3N_q r_e^2}{2} - \alpha r_e. \label{eq:Mext}
\end{equation}

For fixed parameters, Eq.~(\ref{eq:re}) defines $r_e(\alpha)$ and Eq.~(\ref{eq:Mext}) gives $M_{\text{ext}}(\alpha)$, the curve that separates the black-string region ($M > M_{\text{ext}}$, two horizons) from the naked-singularity region ($M < M_{\text{ext}}$, no horizons). The case $M = M_{\text{ext}}$ corresponds to an extremal black string with a single degenerate horizon.

Two special cases are worth noting. If $Q = 0$, Eq.~(\ref{eq:re}) reduces to a quadratic equation for $r_e$, showing that the electric charge is responsible for the quartic nature of the extremal condition. If $\alpha = 0$, the equation remains quartic, indicating that the string-cloud parameter does not simplify the algebraic structure. Thus, the interplay among charge, quintessence, and string cloud leads to a nontrivial horizon structure.

In terms of singularity structure, we must take into account the Kretschmann scalar, which is given in terms of the Riemann tensors $K=R_{\mu\nu\eta\sigma}R^{\mu\nu\eta\sigma}$, taking into account the general solution (\ref{Solution}), we can write
\begin{equation}
\begin{aligned}
  K&=\frac{24}{\ell^{4}}+\frac{48M^{2}}{r^{6}}-\frac{96MQ^{2}}{r^{7}}+\frac{56Q^{4}}{r^{8}}\\&-\frac{12w_{q}\left(3w_{q}-1\right)N_{q}}{\ell^{2}r^{3\left(1+w_{q}\right)}}\\&+\frac{3\left(27w^{4}_{q}+54w^{3}_{q}+51w^{2}_{q}+20w_{q}+4\right)N^{2}_{q}}{r^{6\left(w_{q}+1\right)}}\\&+\frac{24\left(3w_{q}+2\right)\left(w_{q}+1\right)MN_{q}}{r^{\left(6+3w_{q}\right)}}\\&-\frac{12\left(w_{q}+1\right)\left(9w_{q}+4\right)N_{q}Q^{2}}{r^{7+3w_{q}}}\\&+\frac{4\alpha^{2}}{r^{4}}-\frac{8\alpha}{\ell^{2}r^{2}}+\frac{16\alpha M}{r^{5}}-\frac{8\alpha Q^{2}}{r^{6}}+\frac{8\alpha N_{q}}{r^{5+3w_{q}}}.
\end{aligned}
\end{equation}

The case where $\alpha=0$ can be seen in~\cite{Barbosa:2025scy}, now we should note that the presence of $\alpha$ introduces couplings with $M$, $Q$ and $N_q$, enriching the structure of singularities of the black string. 

In addition to the general case, we explicitly compute the Kretschmann scalar for the physically relevant configuration with $w_q=-2/3$:
\begin{equation}
\begin{aligned}
  K&=\frac{24}{\ell^{4}}+\frac{48M^{2}}{r^{6}}-\frac{96MQ^{2}}{r^{7}}+\frac{56Q^{4}}{r^{8}}\\&-\frac{24N_{q}}{\ell^{2}r}+\frac{8N^{2}_{q}}{r^{2}}+\frac{8N_{q}Q^{2}}{r^{5}}\\&-\frac{8\alpha}{\ell^{2}r^{2}}+\frac{8\alpha N_{q}}{r^{3}}+\frac{4\alpha^{2}}{r^{4}}+\frac{16\alpha M}{r^{5}}-\frac{8\alpha Q^{2}}{r^{6}}.
\end{aligned}
\end{equation}

The above expression makes explicit how the curvature invariant receives contributions from the AdS background, the mass and electric charge, and the additional matter sources described by the Kiselev fluid and the string cloud. In particular, the terms with higher inverse powers of $r$ dominate near the central singularity, while the contributions proportional to $N_q$ and $\alpha$ modify the intermediate and asymptotic behavior of the spacetime curvature.

\section{Thermodynamics}\label{Thermodynamics}

Here we discuss thermodynamic aspects of the obtained solution, including the Hawking temperature, heat capacity, and Helmholtz free energy. We present their dependence on the horizon radius, with particular attention to the case $w_q=-2/3$.

\subsection{Temperature}

The Hawking temperature is given by~\cite{Bekenstein:1973ur,Hawking:1975vcx}
\begin{equation}
	T=\left.\frac{f'(r)}{4\pi}\right|_{r=r_h},
\end{equation}
where $r_h$ is the radius of the event horizon. For the solution~\eqref{Solution}, we obtain
\begin{equation}\label{eq:temp_horizon}
	T=\frac{1}{4\pi}\left(
	\frac{2r_h}{\ell^2}
	+\frac{2M}{r_h^2}
	-\frac{2Q^2}{r_h^3}
	+\frac{(3w_q+1)N_q}{r_h^{3w_q+2}}
	\right).
\end{equation}

Using the horizon condition $f(r_h)=0$, the mass parameter $M$ can be expressed as
\begin{equation}\label{eq:mass_horizon}
	M=\frac{r_h}{2}\left(
	\frac{r_h^2}{\ell^2}
	+\frac{Q^2}{r_h^2}
	-\frac{N_q}{r_h^{3w_q+1}}
	-\alpha
	\right).
\end{equation}

Substituting this result into the expression for the Hawking temperature, we obtain
\begin{equation}\label{eq:temperature_general}
	T=\frac{1}{4\pi}\left(
	\frac{3r_h}{\ell^2}
	-\frac{Q^2}{r_h^3}
	+\frac{3w_qN_q}{r_h^{3w_q+2}}
	-\frac{\alpha}{r_h}
	\right).
\end{equation}

	For the case of interest, $w_q=-2/3$, this expression reduces to

\begin{equation}\label{eq:temperature_special}
	T=\frac{1}{4\pi}\left(
	\frac{3r_h}{\ell^2}
	-\frac{Q^2}{r_h^3}
	-2N_q
	-\frac{\alpha}{r_h}
	\right).
\end{equation}

	The quintessence contribution is independent of the horizon radius, while the charge and string-cloud terms scale as $r_h^{-3}$ and $r_h^{-1}$, respectively. For large $r_h$, the leading contribution is $3r_h/(4\pi\ell^2)$.

\begin{figure}[htbp]
	\centering
	\includegraphics[width=\columnwidth]{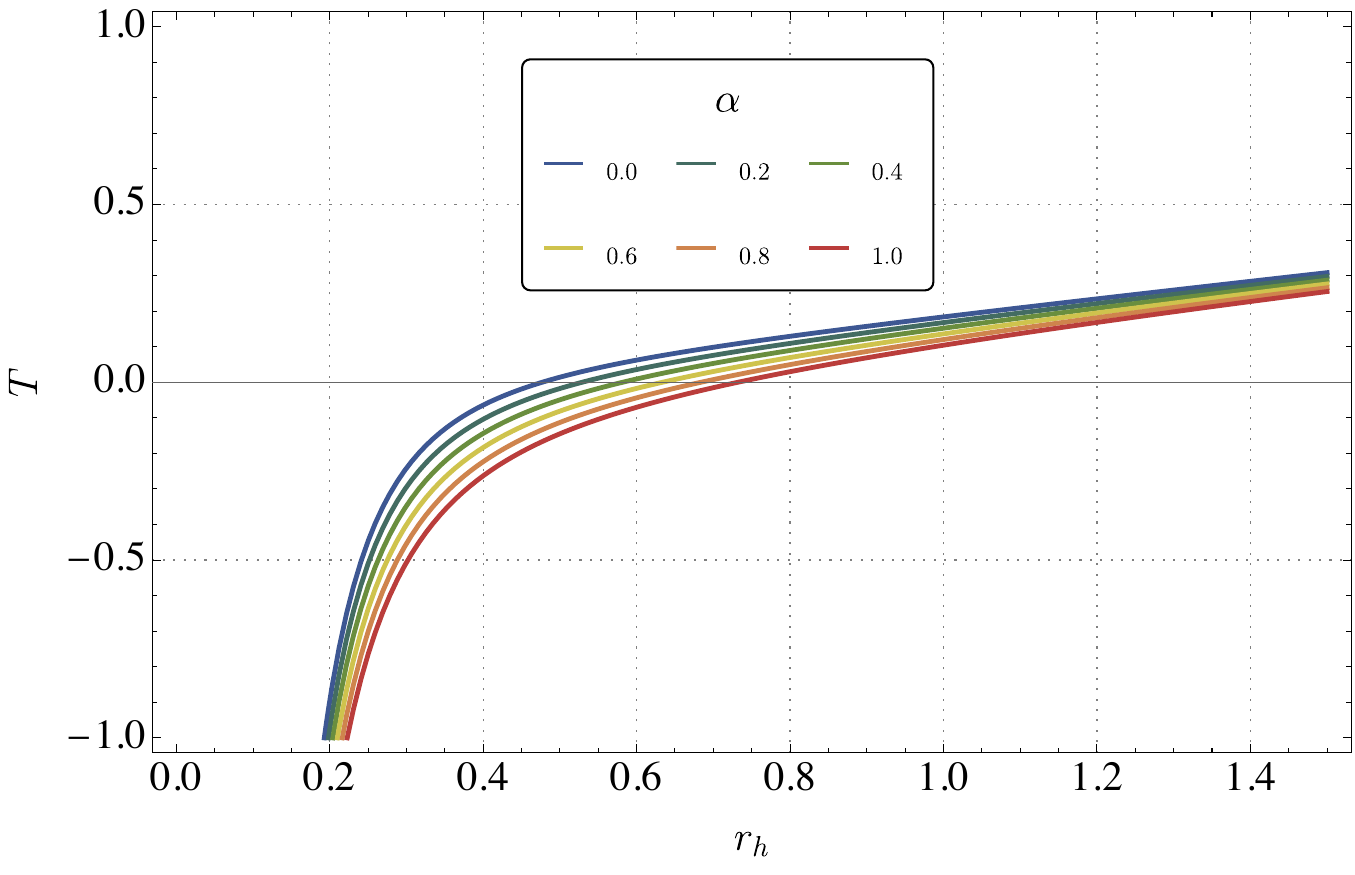}
	\caption{Hawking temperature as a function of the horizon radius for
			$w_q=-2/3$, with $\ell=1$, $Q=0.3$, and $N_q=0.3$. The curves
			correspond to
			$\alpha\in\{0.0,0.2,0.4,0.6,0.8,1.0\}$.}
	\label{fig:temperature}
\end{figure}

\subsection{Heat capacity}
The heat capacity is evaluated at fixed $Q$, $N_q$, $\alpha$, $\ell$, and $w_q$:

\begin{equation}
		C=\left(\frac{\partial M}{\partial T}\right)_{Q,N_q,\alpha,\ell,w_q}.
\end{equation}

Differentiating the expressions for mass and temperature gives
\begin{equation}
	C=\frac{2\pi r^{2}_{h}\left(\frac{r^{2}_{h}}{\ell^{2}}+\frac{w_{q}N_{q}}{r^{3w_{q}+1}_{h}}-\frac{Q^{2}}{3r^{2}_{h}}-\frac{\alpha}{3}\right)}{\left(\frac{r^{2}_{h}}{\ell^{2}}-\frac{\left(3w_{q}+2\right)w_{q}N_{q}}{r^{3w_{q}+1}_{h}}+\frac{Q^{2}}{r^{2}_{h}}+\frac{\alpha}{3}\right)}.
	\label{HeatC}
\end{equation}

	A possible divergence radius, denoted by $r_{\mathrm{div}}$, satisfies

\begin{equation}
		\frac{r_{\mathrm{div}}^2}{\ell^2}
		-\frac{(3w_q+2)w_qN_q}{r_{\mathrm{div}}^{3w_q+1}}
		+\frac{Q^2}{r_{\mathrm{div}}^2}
		+\frac{\alpha}{3}=0,
	\label{eq:critical_general}
\end{equation}

	provided that the numerator of Eq.~\eqref{HeatC} is nonzero at that radius.
 For $w_q=-2/3$, Eq.~\eqref{HeatC} reduces to
\begin{equation}\label{eq:heat_capacity_special}
	C=\frac{2\pi r^{2}_{h}\left(\frac{r^{2}_{h}}{\ell^{2}}-\frac{2N_{q}r_{h}}{3}-\frac{Q^{2}}{3r^{2}_{h}}-\frac{\alpha}{3}\right)}{\left(\frac{r^{2}_{h}}{\ell^{2}}+\frac{Q^{2}}{r^{2}_{h}}+\frac{\alpha}{3}\right)}.
\end{equation}

	In this case, $N_q$ appears only in the numerator. Since
	\begin{equation}
		\frac{r_h^2}{\ell^2}
		+\frac{Q^2}{r_h^2}
		+\frac{\alpha}{3}>0
	\end{equation}
	for $r_h>0$, $\ell^2>0$, and $\alpha\geq0$, the heat capacity has no divergence in the parameter range considered.

\begin{figure}[htbp]
	\centering
	\includegraphics[width=\columnwidth]{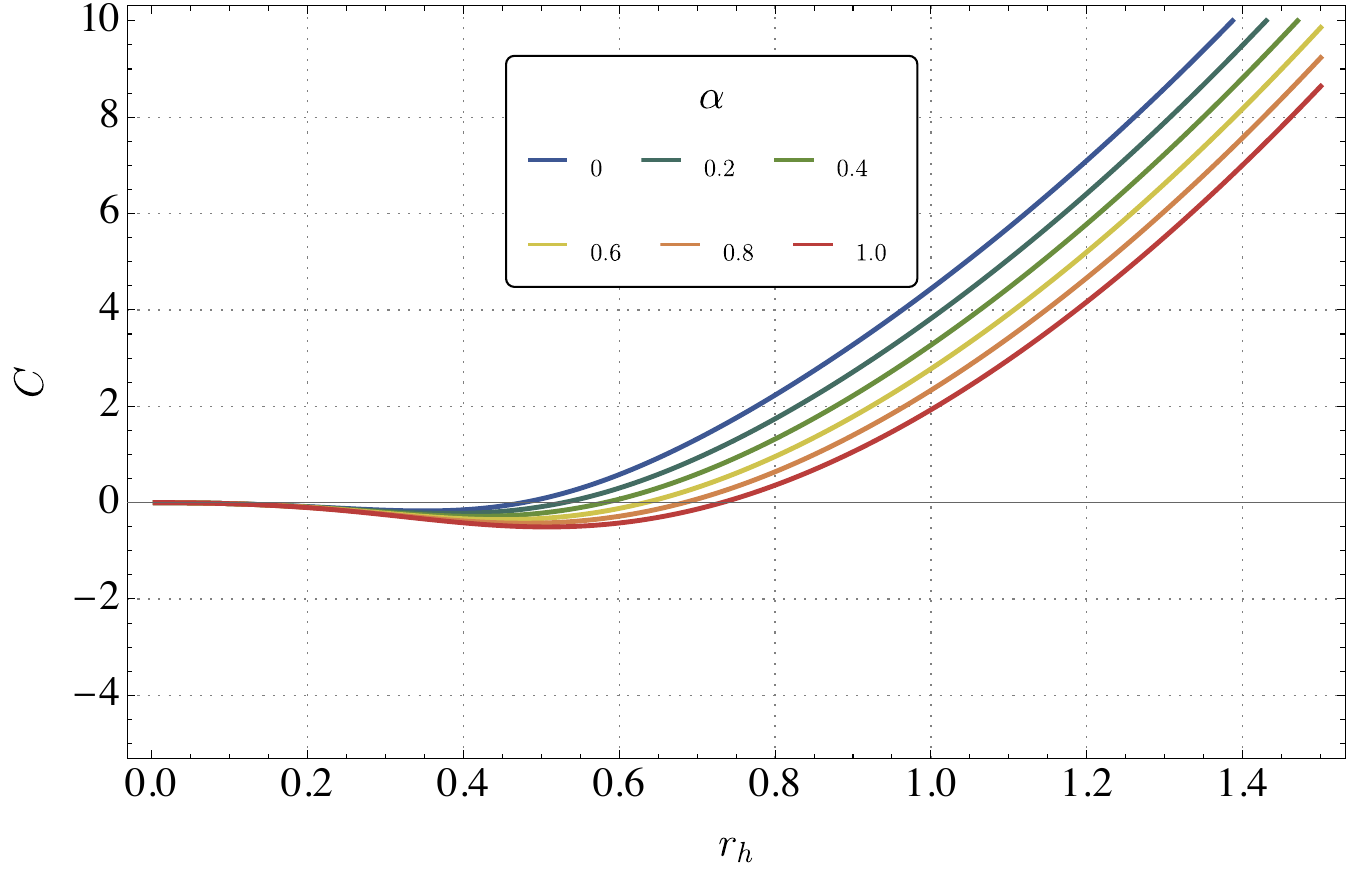}
	\caption{
			Heat capacity as a function of the horizon radius for
			$w_q=-2/3$, with $\ell=1$, $Q=0.3$, and $N_q=0.3$. The curves
			correspond to
			$\alpha\in\{0,0.2,0.4,0.6,0.8,1.0\}$.}
	\label{fig:heat_capacity}
\end{figure}

\subsection{Helmholtz free energy}

We next evaluate the Helmholtz free energy at fixed charge,
\begin{equation}
	F=M-TS,
\end{equation}
evaluated at fixed $Q$, $N_q$, $\alpha$, and $\ell$.
Using $dM=T\,dS$, with the normalization associated with $M$ and a vanishing additive entropy constant, we obtain
\begin{equation}
	S=\pi r_h^2.
	\label{eq:entropy}
\end{equation}

Using this result together with Eqs.~\eqref{eq:mass_horizon} and~\eqref{eq:temperature_general}, the free energy for general $w_q$ reads
\begin{equation}\label{eq:free_energy_general}
	F=\frac{1}{4}\left[
	-\frac{r_h^3}{\ell^2}
	-\frac{(3w_q+2)N_q}{r_h^{3w_q}}
	+\frac{3Q^2}{r_h}
	-\alpha r_h
	\right].
\end{equation}

For $w_q=-2/3$, the coefficient $(3w_q+2)$ vanishes and the explicit dependence on $N_q$ drops out:
\begin{equation}\label{eq:free_energy_special}
	F=\frac{1}{4}\left[
	-\frac{r_h^3}{\ell^2}
	+\frac{3Q^2}{r_h}
	-\alpha r_h
	\right].
\end{equation}

Differentiating Eq.~\eqref{eq:free_energy_special} with respect to $r_h$ gives
\begin{equation}
		\frac{dF}{dr_h}
		=
		-\frac{3}{4}\left[
		\frac{r_h^2}{\ell^2}
		+\frac{Q^2}{r_h^2}
		+\frac{\alpha}{3}
		\right].
	\label{eq:dFdr_special}
\end{equation}
For $Q\neq0$, $\alpha>0$, and $\ell>0$, the bracket is positive for all $r_h>0$, so $F$ is strictly decreasing. Since $F\to+\infty$ as $r_h\to0^{+}$ and $F\to-\infty$ as $r_h\to+\infty$, $F$ has exactly one positive zero $r_F$, defined by $F(r_F)=0$:
\begin{equation}
		r_F^4+\alpha\ell^2r_F^2-3\ell^2Q^2=0.
	\label{eq:free_energy_zero}
\end{equation}

For $Q\neq0$, the positive solution is
\begin{equation}
	r_F=
	\sqrt{
		\frac{
			-\alpha\ell^2+\sqrt{\alpha^2\ell^4+12\ell^2Q^2}
		}{2}
	}.
	\label{eq:free_energy_radius}
\end{equation}
The other root for $r_F^2$ is negative. This expression remains valid at $\alpha=0$.

The sign of $F$ then follows from its monotonicity,
\begin{equation}
	F(r_h)
	\begin{cases}
		> 0, & r_h < r_F, \\[2pt]
		= 0, & r_h = r_F, \\[2pt]
		< 0, & r_h > r_F,
	\end{cases}
	\label{eq:F_inequalities}
\end{equation}
so that the three cases $F>0$, $F=0$, and $F<0$ correspond respectively to $M>TS$, $M=TS$, and $M<TS$.

\begin{figure}[htbp]
	\centering
	\includegraphics[width=\columnwidth]{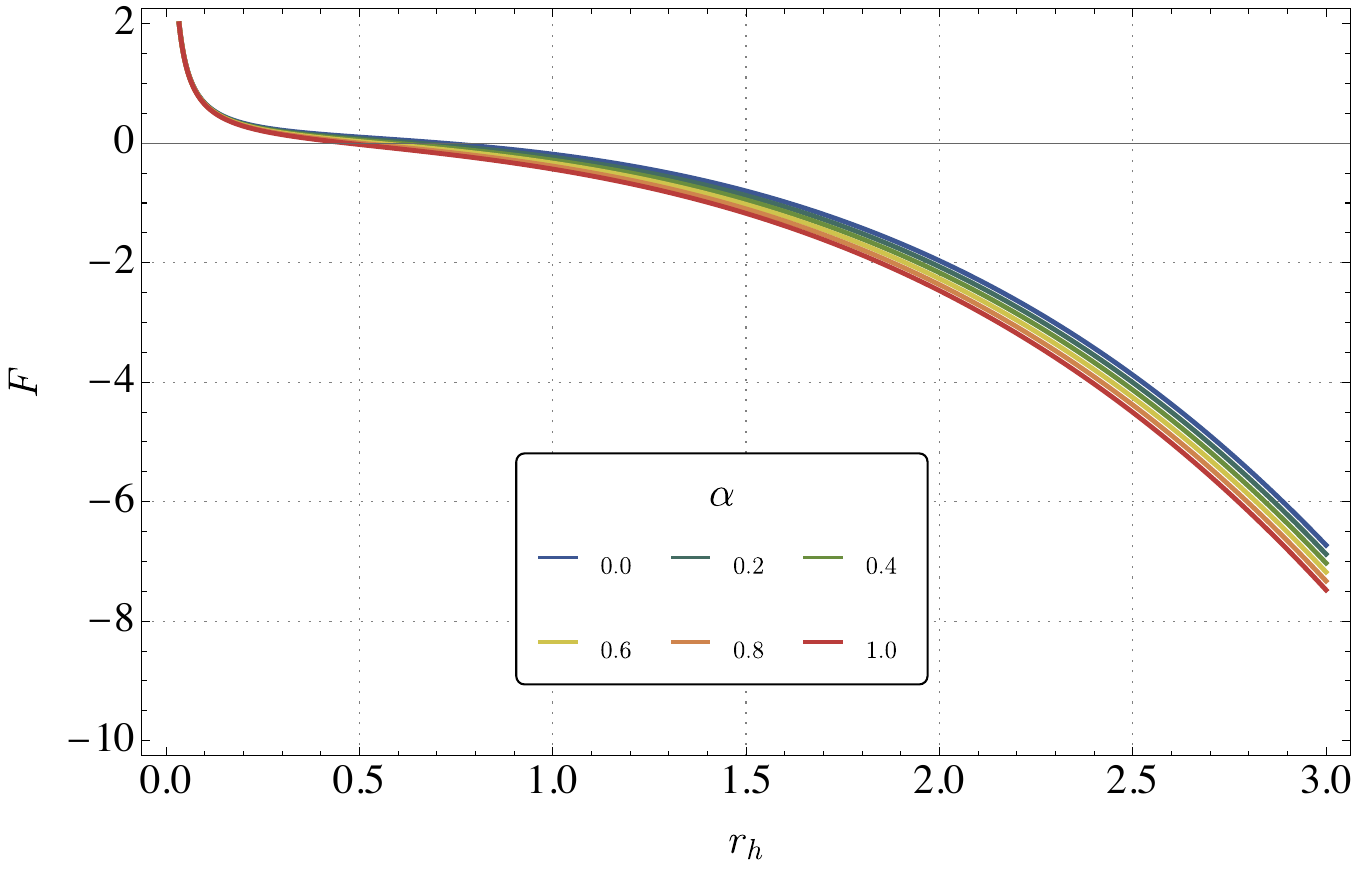}
	\caption{
			Helmholtz free energy as a function of the horizon radius for
			$w_q=-2/3$, with $\ell=1$ and $Q=0.3$. The curves correspond
			to $\alpha\in\{0.0,0.2,0.4,0.6,0.8,1.0\}$.}
	\label{fig:free_energy}
\end{figure}

\section{Photon cylinder}\label{Photon_cylinder}

	In this section, we investigate whether the black string admits null geodesics at constant radius outside the outer horizon. These trajectories lie on a cylinder \(r=r_c\) and are circular when the longitudinal momentum vanishes, or helical when both the angular and longitudinal momenta are nonzero.

For null geodesics, the motion follows from the Lagrangian
\begin{equation}
	\mathcal{L}=\frac{1}{2}g_{\mu\nu}\dot{x}^{\mu}\dot{x}^{\nu},
	\label{eqL}
\end{equation}
where the overdot denotes differentiation with respect to an affine parameter. Setting \(\mathcal{L}=0\), the line element gives
\begin{equation}
	2\mathcal{L}
	=-f(r)\dot{t}^{2}
	+\frac{\dot{r}^{2}}{f(r)}
	+r^{2}\dot{\varphi}^{2}
	+\frac{r^{2}}{\ell^{2}}\dot{z}^{2}=0.
\end{equation}
The Killing vectors \(\partial_t\), \(\partial_\varphi\), and \(\partial_z\) yield the conserved quantities
\begin{equation}
	E=f(r)\dot{t},\qquad
	L_\varphi=r^2\dot{\varphi},\qquad
	P_z=\frac{r^2}{\ell^2}\dot{z}.
\end{equation}
Using these expressions, the radial equation can be written as
\begin{equation}
	\dot{r}^{2}=E^{2}-V_{\mathrm{eff}}(r),
\end{equation}
with the effective potential
\begin{equation}
	V_{\mathrm{eff}}(r)
	=\frac{\lambda^{2}f(r)}{r^{2}},
	\qquad
	\lambda^{2}=L_{\varphi}^{2}+\ell^{2}P_z^{2}.
\end{equation}

	For \(\lambda^2>0\), a null geodesic at constant radius requires both vanishing radial velocity and vanishing radial acceleration. These conditions read
	\begin{equation}
		E^2=V_{\mathrm{eff}}(r_c),
		\qquad
		V_{\mathrm{eff}}'(r_c)=0.
		\label{eq:photon_conditions}
	\end{equation}
	The second condition identifies the stationary radii of the potential, while the first requires a compatible real energy. In particular, a stationary point with \(V_{\mathrm{eff}}(r_c)<0\) does not describe a constant-radius null geodesic.
	
	Substituting the metric function of Eq.~\eqref{Solution} into the effective potential and differentiating with respect to \(r\), we obtain
	\begin{equation}
		V_{\mathrm{eff}}'(r)
		=\frac{\lambda^2}{r^5}
		\left[
		6Mr-4Q^2
		+3(w_q+1)N_qr^{1-3w_q}
		+2\alpha r^2
		\right].
		\label{eq:photon_potential_derivative}
	\end{equation}
	To examine the sign of this expression outside the outer horizon, we use \(f(r_h)=0\) to eliminate the charge parameter:
	\begin{equation}
		Q^2
		=2Mr_h-\frac{r_h^4}{\ell^2}
		+N_qr_h^{1-3w_q}
		+\alpha r_h^2.
		\label{eq:charge_horizon}
	\end{equation}
	After this substitution, the terms in
	Eq.~\eqref{eq:photon_potential_derivative} can be separated into radial differences and a contribution evaluated at the horizon. To identify the latter, we evaluate \(f'(r_h)\) using Eq.~\eqref{eq:charge_horizon}, which gives
	\begin{equation}
		\begin{aligned}
			r_h^3f'(r_h)
			={}&\frac{4r_h^4}{\ell^2}-2Mr_h\\
			&+(3w_q-1)N_qr_h^{1-3w_q}
			-2\alpha r_h^2.
		\end{aligned}
		\label{eq:photon_horizon}
	\end{equation}
	The derivative of the effective potential can then be written as
	\begin{equation}
		\begin{aligned}
			\frac{r^5}{\lambda^2}V_{\mathrm{eff}}'(r)
			={}&r_h^3f'(r_h)+6M(r-r_h)\\
			&+2\alpha(r^2-r_h^2)\\
			&+3(w_q+1)N_q
			\left(r^{1-3w_q}-r_h^{1-3w_q}\right).
		\end{aligned}
		\label{eq:photon_exterior_identity}
	\end{equation}
	
	We consider \(M>0\), \(N_q>0\), \(\alpha\geq0\), and
	\(-1\leq w_q\leq-1/3\), assuming the existence of an outer horizon bounding a static exterior. Since \(f(r_h)=0\) and \(f(r)>0\) immediately outside the horizon, \(f'(r_h)\geq0\). This derivative is positive for a simple outer horizon and vanishes for a degenerate horizon.
	
	For \(r>r_h\), the mass contribution in
	Eq.~\eqref{eq:photon_exterior_identity} is strictly positive, while the string-cloud contribution is nonnegative. The Kiselev contribution is also nonnegative because \(w_q+1\geq0\), \(N_q>0\), and \(1-3w_q>0\), so that \(r^{1-3w_q}>r_h^{1-3w_q}\). Consequently,
	\begin{equation}
		V_{\mathrm{eff}}'(r)>0,
		\qquad r>r_h.
		\label{eq:no_exterior_photon}
	\end{equation}
	The effective potential therefore has no stationary point in the exterior region. Under the stated assumptions, the solution admits no exterior photon cylinder, including when the outer horizon is degenerate.
	
\begin{figure}[htbp]
	\centering
	\includegraphics[width=\columnwidth]{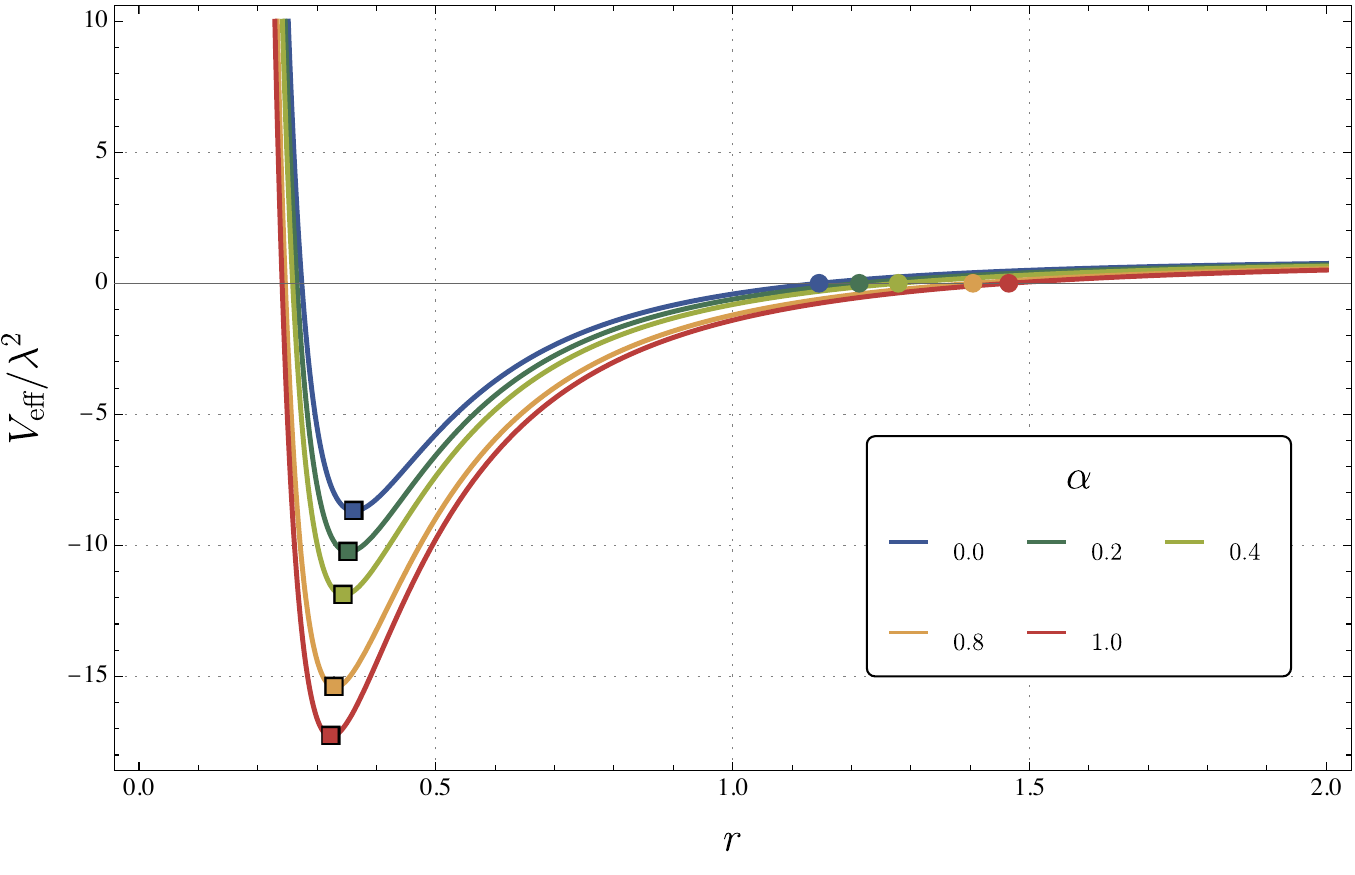}
	\caption{
			Effective potential \(V_{\mathrm{eff}}/\lambda^2\)
			as a function of \(r\), with \(w_q=-2/3\),
			\(\ell=1\), \(M=0.9\), \(Q=0.7\), and \(N_q=0.1\).
			The curves correspond to
			\(\alpha\in\{0.0,0.2,0.4,0.8,1.0\}\).
			Filled circles mark the outer event horizons \(r_h\),
			and squares mark the stationary radii \(r_c\),
			where \(V_{\mathrm{eff}}'(r_c)=0\).
			The minima lie inside the outer horizons and have
			negative potential values, while the potential
			increases monotonically in the exterior region.
	}
	\label{fig:photon_cylinder}
\end{figure}

	Figure~\ref{fig:photon_cylinder} illustrates this behavior for representative parameter values and different values of \(\alpha\). The circles mark the outer event horizons, where the effective potential vanishes, while the squares indicate the stationary radii \(r_c\). For the curves shown, these stationary points are minima located inside the outer horizons and have negative potential values. They therefore do not satisfy the energy condition in Eq.~\eqref{eq:photon_conditions}. Outside each event horizon, the potential increases monotonically, in agreement with the analytical result.

\section{Discussion and conclusions}\label{Discussion_and_conclusions}

	We obtained an exact static charged black-string solution in Einstein--Maxwell theory with a negative cosmological constant, a Kiselev fluid, and a string cloud. The metric was derived for general \(w_q\), while the horizon and thermodynamic analyses focused on \(w_q=-2/3\). The limits \(\alpha=0\) and \(Q=0\) recover the related charged and neutral configurations discussed in the appendix.
	
	The energy-condition analysis distinguished the contributions of each matter sector and included the cosmological constant as an effective source. In the adopted convention, positive Kiselev energy density requires \(N_q>0\), while the string-cloud density is nonnegative for \(\alpha\geq0\). We also examined the inner and outer horizons and obtained the conditions for an extremal configuration. The Kretschmann scalar shows that the curvature singularity at \(r=0\) persists.
	
	The thermodynamic analysis provided expressions for the Hawking temperature, heat capacity, and Helmholtz free energy. For \(w_q=-2/3\) and \(\alpha\geq0\), the heat-capacity denominator is strictly positive for \(r_h>0\), so no divergence occurs. In this case, the explicit dependence on \(N_q\) drops out of the Helmholtz free energy. For nonzero charge, \(F\) decreases monotonically with \(r_h\) and has a single positive zero \(r_F\). This zero alone does not establish a phase transition or global thermodynamic stability.
	
	Finally, we examined constant-radius null geodesics for general \(w_q\). Using \(f(r_h)=0\) to eliminate \(Q^2\), we showed that \(V_{\mathrm{eff}}'(r)>0\) outside the outer horizon for \(M>0\), \(N_q>0\), \(\alpha\geq0\), and \(-1\leq w_q\leq-1/3\), assuming a static exterior. Thus, no exterior photon cylinder exists under these assumptions, including for a degenerate outer horizon. The plotted stationary points lie inside the outer horizons and have negative potential values, so they do not represent constant-radius null geodesics.
	
	Possible extensions include rotating configurations and the study of scalar perturbations and quasinormal modes.

\section{Acknowledgements}\label{Acknowledgements}

	The authors thank the anonymous referee for the valuable suggestions, which greatly enriched the manuscript.  L.G.B. acknowledges the financial support of the Coordenação de Aperfeiçoamento de Pessoal de Nível Superior (CAPES), Brazil (Finance Code 001). F.M.S. would like to thank CNPq for financial support under research project No. 403007/2024-0 and research fellowship No. 201145/2025-1.

\appendix

\section{Comparison with related black-string solutions}
\label{app:comparison}

	Table~\ref{tab:static_comparison} compares the solution obtained in this work with related static black-string solutions in the literature. The metric functions are presented in a common notation, alongside their matter sources and the analyses performed in each reference.

	\begin{table*}[h]
		\centering
		\caption{
				Comparison of static black-string solutions in a common geometric notation, with $\Lambda=-3/\ell^2$. The sign conventions adopted for the solutions follow those used in the present work. The last column summarizes selected analyses and extensions presented in each reference.}
		\label{tab:static_comparison}
		
		\small
		\setlength{\tabcolsep}{5pt}
		\renewcommand{\arraystretch}{1.35}
		
		\begin{tabularx}{\textwidth}{@{}
				>{\raggedright\arraybackslash}p{0.18\textwidth}
				>{\raggedright\arraybackslash}p{0.18\textwidth}
				>{\raggedright\arraybackslash}p{0.26\textwidth}
				>{\raggedright\arraybackslash}X
				@{}}
			\hline
			\hline
			\textbf{Reference}
			&
			\textbf{Matter sectors}
			&
			\textbf{Metric function \(f(r)\)}
			&
			\textbf{Analyses}
			\\
			\hline
			
			Ali et al.~\cite{Ali:2019mxs}
			&
			Kiselev fluid
			&
			$\displaystyle
			\frac{r^2}{\ell^2}-\frac{2M}{r}
			-\frac{N_q}{r^{3w_q+1}}$
			&
			Fluid properties, curvature, horizons, rotating extension,
			conserved quantities, and thermodynamics.
			\\
			\hline
			
			Barbosa et al.~\cite{Barbosa:2025scy}
			&
			Maxwell field and Kiselev fluid
			&
			$\displaystyle
			\frac{r^2}{\ell^2}-\frac{2M}{r}
			+\frac{Q^2}{r^2}-\frac{N_q}{r^{3w_q+1}}$
			&
			Energy conditions, curvature, horizons, rotating extension,
			closed timelike curves, conserved charges, scalar tunneling,
			and thermodynamics.
			\\
			\hline
			
			Deglmann et al.~\cite{Deglmann:2025mcl}
			&
			String cloud and Kiselev fluid
			&
			$\displaystyle
			\frac{r^2}{\ell^2}-\frac{2M}{r}
			-\frac{N_q}{r^{3w_q+1}}-\alpha$
			&
			Fluid density and pressures, parameter estimates,
			horizon formation, and near-horizon Klein--Gordon solutions.
			\\
			\hline
			
			Present work
			&
			Maxwell field, Kiselev fluid, and string cloud
			&
			$\displaystyle
			\frac{r^2}{\ell^2}-\frac{2M}{r}
			+\frac{Q^2}{r^2}-\frac{N_q}{r^{3w_q+1}}-\alpha$
			&
			Energy conditions, Kretschmann scalar, horizons and
			extremality, temperature, heat capacity, Helmholtz free
			energy, and circular null geodesics.
			\\
			\hline
			\hline
		\end{tabularx}
	\end{table*}

\section*{Generative AI declaration}
During the preparation of this work, the authors used the free version of ChatGPT (OpenAI) solely for language polishing and improving the clarity of the manuscript. The authors reviewed and edited the output as needed and take full responsibility for the content of the published article.

\bibliographystyle{unsrturl}
\bibliography{sample}

\end{document}